\documentclass[aps,prf,reprint,onecolumn,superscriptaddress, longbibliography, 12pt]{revtex4-1}

\usepackage{graphicx}
\usepackage{xcolor}

\usepackage{dcolumn}   
\usepackage{bm}        
\usepackage{amssymb}   
\usepackage{amsmath}
\usepackage[normalem]{ulem}

\usepackage{siunitx}
	 \DeclareSIUnit{\wtpercent}{wt\%}
	\DeclareSIUnit\eV{\eV}
    \DeclareSIUnit\bar{bar}
	
	\usepackage{cleveref}
\newcommand{\MassFracW}{w}
\newcommand{\VapConcW}{c}
\newcommand{\VapConcWVLE}{c_\mathrm{VLE}}
\newcommand{\VapConcWInfty}{c_\infty}
\newcommand{\MassTransW}{j}

\usepackage{booktabs}

\begin{document}

\title{Role of surfactants on droplet formation in piezoacoustic inkjet printing across microsecond-to-second timescales}

\author{Maaike Rump}
\affiliation{Physics of Fluids group, Max Planck Center Twente for Complex Fluid Dynamics, and J. M. Burgers Centre for Fluid Dynamics, University of Twente, 7500AE Enschede, The Netherlands}
\author{Christian Diddens}
\affiliation{Physics of Fluids group, Max Planck Center Twente for Complex Fluid Dynamics, and J. M. Burgers Centre for Fluid Dynamics, University of Twente, 7500AE Enschede, The Netherlands}
\author{Uddalok Sen}
\affiliation{Physical Chemistry and Soft Matter, Wageningen University and Research, 6708WE Wageningen, The Netherlands}
\author{Michel Versluis}
\affiliation{Physics of Fluids group, Max Planck Center Twente for Complex Fluid Dynamics, and J. M. Burgers Centre for Fluid Dynamics, University of Twente, 7500AE Enschede, The Netherlands}
\author{Detlef Lohse}
\affiliation{Physics of Fluids group, Max Planck Center Twente for Complex Fluid Dynamics, and J. M. Burgers Centre for Fluid Dynamics, University of Twente, 7500AE Enschede, The Netherlands}
\author{Tim Segers}
\affiliation{BIOS / Lab on a Chip group, Max Planck Center Twente for Complex Fluid Dynamics, MESA+ Institute for Nanotechnology, University of Twente, 7500AE Enschede, The Netherlands}

\date{\today}

\maketitle


\section*{Abstract}
In piezoacoustic drop-on-demand inkjet printing, a single droplet is produced for each piezo driving pulse. This droplet is typically multicomponent, including surfactants to control the spreading and drying of the droplet on the substrate. However, the role of these surfactants on the droplet formation process remains rather elusive. Surfactant concentration gradients may manifest across microsecond-to-second timescales, spanning both the rapid ejection of ink from the nozzle exit and the comparatively slower idling timescale governing the firing of successive droplets. In the present work, we study the influence of surfactants on droplet formation across 6~orders of magnitude in time. To this end, we visualize the microsecond droplet formation process using stroboscopic 8-ns laser-induced fluorescence microscopy while we vary the nozzle idle time. Our results show that increasing the idle time up to $\mathcal{O}$(1)~s affects only the break-up dynamics of the inkjet but not its velocity. By contrast, for idle times $>$~$\mathcal{O}$(1)~s, both the break-up dynamics are altered and the velocity of the inkjet increases. We show that the increased velocity results from a decreased surface tension of the ejected droplet, which we extracted from the observed shape oscillations of the jetted droplets in flight. The measured decrease in surface tension is surprising as the \SI{}{\micro \second} timescale of droplet formation is much faster than the typical ms-to-s timescale of surfactant adsorption. By varying the bulk surfactant concentration, we show that the fast decrease in surface tension results from a local surfactant concentration increase to more than 200~times the CMC.  Numerical simulations then show that the evaporation-driven increased surfactant concentration present at the nozzle exit fully coats the surface of the droplet during its ejection. Altogether, our results suggest that a local high concentration of surfactant allows for surfactant adsorption to the interface of an inkjet at the  \SI{}{\micro \second}-to-ms timescale, which is much faster than the typical ms-to-s timescale associated to surfactant adsorption.

\vspace{1 cm}


\section{Introduction}

Drop-on-demand inkjet printing enables precise, non-contact deposition of picoliter droplets~\cite{Wijshoff2010,Derby2010,Basaran2013,Hoath2016,Lohse2022}, accommodating liquids with a diverse range of physical properties. Beyond traditional graphics printing on paper, inkjet technology finds application in the fabrication of electroluminescent displays~\cite{Shimoda2003,Wei2022}, electronic circuits~\cite{Sirringhaus2000,Majee2016,Majee2017}, and biomaterials~\cite{Villar2013,Daly2015,Simaite2016}.

In inkjet printing, surfactants are commonly included in the ink formulation to regulate the spreading and drying behavior of the deposited droplets~\cite{Wijshoff2018}. By reducing surface tension~\cite{Squires2020}, surfactants promote droplet spreading~\cite{Bera2018,Hack2021a}. Additionally, they serve to mitigate the so-called coffee-stain effect~\cite{kim-2016-prl,Marin2016,vGaalen2021} during droplet drying, ensuring a uniform distribution of deposited pigment particles~\cite{Li2020b}. Besides influencing spreading and drying dynamics, surfactants may also affect the fast ejection and formation process of droplets prior to substrate contact. 

\begin{figure*}[tb]
	\includegraphics[width=1\textwidth]{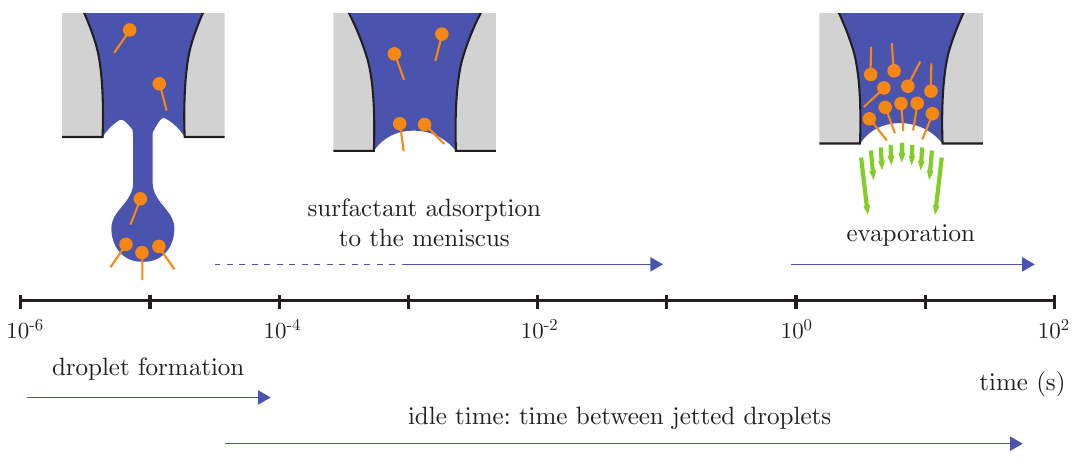}
	\caption{Inkjet printing operates across microsecond-to-second timescales. The timescale of droplet formation ($10^{-5} - 10^{-4}$~s; left)  is generally considered to be too fast for surfactant adsorption to the newly formed interface. As such, a non-homogenous distribution of surfactants is expected on the inkjet. The timescale that follows is that of the interval between the firing of successive droplets ($10^{-5} - 10^{-1}$ s; middle). This idle time can extend to timescales where evaporation starts to play a role ($10^{0} - 10^{2}$ s; right)~\cite{RumpDrying}, which may result in surfactant accumulation at the nozzle exit. }
	\label{fig:Surfintro}
\end{figure*}

Inkjet printing operates across multiple timescales, each characterized by distinct flow features~\cite{Hoath2016}, as illustrated in Fig.~\ref{fig:Surfintro}. The shortest timescale ($10^{-5} - 10^{-4}$ s; left) is associated with droplet formation. The timescale that follows is the timescale involved in the interval between the firing of successive droplets ($10^{-5} - 10^{-1}$ s; middle). This idle time can extend to timescales where evaporation starts to play a role ($\approx 10^{0} - 10^{2}$ s; right in Fig.~\ref{fig:Surfintro})~\cite{RumpDrying}. Evaporation of  liquid from the nozzle exit may lead to a local increase of the surfactant concentration in the nozzle, potentially exceeding the CMC by orders of magnitude. 

As detailed by Manikantan and Squires in their seminal review article~\cite{Squires2020}, surfactant transport in the jetted ink is governed by three key physical processes: advection of surfactant molecules by the flow,  diffusion across concentration gradients, and surface phenomena, including adsorption, desorption, and diffusion at the interface~\cite{Squires2020}. Each of the aforementioned processes is characterized by its own associated time scale. 

The competition between advective and diffusive transport can be quantified by the P\'{e}clet number ($\mathrm{Pe}$), which is defined as the ratio of the timescale for diffusion to that of advection. The advection timescale in inkjet printing (inkjet radius / inkjet velocity) is of the order of \SI{10}{\micro \second}. The characteristic timescale for bulk transport via diffusion can be expressed as: 
\begin{equation} \label{Eq:tauD}
\tau_D \simeq \frac{\Gamma_{eq}^2} {D c_b^2},
\end{equation}
 with $\Gamma_{eq}$ the equilibrium surface concentration, $D$ the bulk diffusion coefficient, and $c_b$ the bulk concentration~\cite{Ward1946,Pan1998, He2015,Antonopoulou2021}. For the surfactant used in this work (Triton X-100), $\tau_D$ is 0.2~s at a $c_b$ that equals the critical micelle concentration (CMC)~\cite{Horozov2000}. The resulting $\mathrm{Pe}$ of 20,000 indicates that in inkjet printing, typically, surfactant transport by advection dominates that by diffusion. 

The surface diffusion coefficient of surfactants is reported to be similar to that in the bulk~\cite{vanGaalen2020}. Consequently, also surfactant transport on the surface is expected to be dominated by advection instead of diffusion.
 Therefore, a non-homogenous distribution of surfactant is expected on the inkjet during its ejection from the nozzle~\cite{Jiang2015}. The difference in surface tension along the interface can result in Marangoni flows, which have been shown to delay both the thinning rate of a surfactant covered liquid column and its break-up~\cite{Craster2002,Kovalchuk2016,Kovalchuk2018,Kamat2020,Martinez2020,Antonopoulou2021,Sijs2021,Kamat2018}. Notably, \citet{Antonopoulou2021} observed in their numerical simulations that the surfactants remain at the front of the liquid column; therefore the pinch-off from the nozzle is not influenced by the surfactants initially present on the meniscus in the nozzle, but only the breakup of the liquid tail into smaller droplets. Only recently, studies have aimed to incorporate the role of surface rheology of a surfactant monolayer on the thinning of a liquid jet ~\cite{Wee2021,Ponce2017}.

In the limit of an infinitely fast surfactant supply to the subsurface of a clean interface, the timescale for surfactants to alter the surface tension is rate limited and governed by adsorption and desorption rates that depend both on the surface and bulk concentrations~\cite{Squires2008}. These rate constants are typically too fast to measure accurately and, as a result, remain largely unknown~\cite{He2015}. 
As adsorption is faster than diffusion, the Damk\"{o}hler number ($\mathrm{Da}$) -- the ratio of the diffusion timescale $\tau_D$ to the kinetic adsorption timescale -- is generally expected to be much larger than 1 in inkjet printing.
The adsorption and desorption rate constants govern the relation between the surface concentration and the bulk  concentration, as described by models like the Langmuir isotherm. Through the Gibbs equation, a surface equation of state can be obtained that links surface tension to bulk surfactant concentration, such as the Langmuir-Gibbs equation of state~\cite{Martinez2020b}. However, these equations are only valid below the CMC and under equilibrium conditions. Despite numerous studies, adsorption and desorption behaviors, particularly at surfactant concentrations above the CMC and on short timescales far from equilibrium, remain largely unclear~\cite{He2015}.

In summary, in inkjet printing, surfactant concentration gradients may manifest across microsecond-to-second timescales, thereby potentially affecting droplet formation. 
The physical transport mechanisms by which surfactants affect droplet formation across these timescales are largely intertwined. In addition, knowledge of surfactant  parameters such as (surface) diffusion coefficients, adsorption and desorption rate constants, also at surfactant concentrations $>$ CMC, and the role of micelles in surfactant transport and kinetics remain largely elusive~\cite{Song2011}. Therefore, in this work, we follow an experimental approach.
 We investigate the effect of both a slow and a fast surfactant on droplet formation across timescales spanning 6~orders of magnitude. To this end, we stroboscopically visualize the droplet formation process using 8-ns laser-induced fluorescence illumination pulses. Additionally, we visualize the shape oscillation of the droplets in flight, i.e. after pinch-off, to extract the surface tension of the jetted droplet~\cite{Staat2017,Yang2014,Bzdek2020,Hoath2015}. Further, we employ numerical simulations including evaporation and surfactant transport to unravel the physical principles underlying the experimental observations. 

The paper is organized as follows. We start by describing the experimental methods and analysis in Section~\ref{S2}, followed by the numerical method in Section~\ref{sec:num}. Section~\ref{S4} contains the results and discussion. The paper ends with the conclusions and an outlook in Section~\ref{S5}.

\section{Experimental methods} \label{S2}

\subsection{Printhead and ink}
Autodrop Pipettes (nozzle diameter: \SI{70}{\, \micro\meter} and \SI{50}{\, \micro\meter}) from Microdrop Technologies GmbH (AD-K-501 and AD-H-501) were used as printheads (Fig.~\ref{fig:ExpSetups}a). They consist of a cylindrical piezoacoustic transducer bonded around a glass capillary connected to a fluid reservoir. More details about the printhead can be found in references~\cite{Fraters2021,Rump2022,sen-2021-jfm}. The printhead is set to drive the meniscus in a push-pull motion, thereby first ejecting a liquid filament and then retracting the meniscus, a process that leads to pinch-off. The surfactant solutions were supplied from a pressure controlled reservoir set to a constant underpressure of \SI{8}{\, \milli\bar} to prevent the liquid from dripping out of the nozzle due to gravity.

Two different surfactants were used: the non-ionic Gemini Dynol 607 (Air Products, \SI{342}{\, \gram/\mole}) and the more commonly used non-ionic Triton X-100 (Sigma-Aldrich, \SI{647}{\, \gram/\mole}), which was also used by \citet{Antonopoulou2021}. The CMCs of these aqueous solutions are \SI{2.92}{\, \mol/\meter\cubed} (\SI{0.1}{\wtpercent}) (see supplementary information Fig.~S1), and \SI{0.22}{\, \mol/\meter\cubed} (\SI{0.014}{\, \wtpercent})~\cite{Tiller1984}, respectively. Dynol, being a Gemini surfactant, entails that it has multiple hydrophilic heads and multiple hydrophobic tails, linked by spacers~\cite{menger-1991-jacs, penkina-2020-aipconfproc}. The close packing of the hydrophobic tails in a monolayer of Gemini surfactants results in a more pronounced reduction of the surface tension for Gemini surfactants as compared to conventional surfactants~\cite{Kamal2016}. Dynamic surface tension measurements show that after \SI{50}{\, \milli\second}, Dynol has reduced the surface tension of water to \SI{40}{\, \milli\newton/\meter} (from $\approx$ \SI{70}{\, \milli\newton/\meter}, see supplementary information Fig.~S2) while Triton is much slower, reducing the surface tension to only \SI{60}{\, \milli\newton/\meter} after \SI{50}{\, \milli\second}, as shown in~\cite{Wasekar2002}.

\subsection{Imaging setups}

\begin{figure}[h!]
	\centering
	\includegraphics[width=.6\columnwidth]{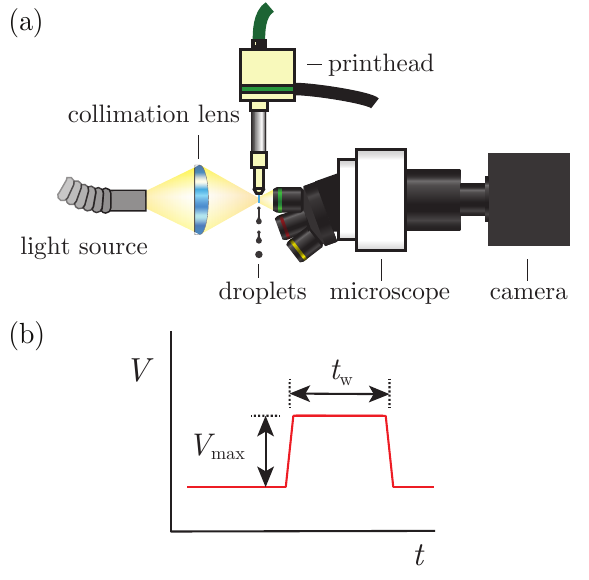}
	\caption{(a) Schematic of the imaging setup that either used stroboscopic or continuous illumination in combination with a slow-speed CCD camera or an ultra-high-speed camera, respectively. (b) Sketch of the driving pulse where $V_{\max}$ is the pulse amplitude and $t_\mathrm{w}$ the pulse width.}
	\label{fig:ExpSetups}
\end{figure}

Two imaging methods were used to perform the measurements. The first method uses stroboscopic imaging, where the nozzle was illuminated by incoherent \SI{8}{\nano\second} light pulses. The illumination pulses were generated by a laser-induced fluorescence (iLIF) system~\cite{vanderbos2011ilif}. The iLIF system consists of a pulsed laser (Quantel EverGreen, dual cavity Nd:YAG, 532~nm, 7~ns) that excites a fluorescent plate embedded in a highly efficient diffuser (Lavision, part nos. 1108417 and 1003144). The emitted light was coupled into an optical fiber that illuminated the printhead via a collimating lens, see Fig.~\ref{fig:ExpSetups}a. The employed microscope (BX-FM Olympus) had a 5$\times$ objective (LMPLFLN 5$\times$, NA of 0.13), a tube lens (U-TLU), and a high-resolution CCD camera (Lumenera, Lw135m, 1392$~\times~$1040~pixels, \SI{4.65}{\, \micro\meter}/pixel size). The resulting optical resolution was \SI{0.93}{\, \micro\meter}/pixel. The images captured by the camera were saved on a personal computer by in-house and custom-made software programmed in the graphical programming language \textsc{Labview} (National Instruments).

The second imaging method comprises ultra-high-speed imaging at 10 million frames per second. The high-speed camera (Shimadzu HPV-X2, 400$~\times~$250~pixels, \SI{32}{\, \micro\meter} square pixels) captured frames illuminated by a continuous fiber-optic light source (LS-M352, Sumita) using the same setup as before (Fig.~\ref{fig:ExpSetups}a). High-speed imaging was employed to record the droplet formation process and the oscillation modes of the jetted droplets, to measure their surface tension. Both a 5$\times$ and a 10$\times$ objective (LMPLFLN, NA of 0.13 and 0.25, respectively) were used, with a resulting imaging resolution of \SI{6.64}{\, \micro\meter/pixel} and \SI{3.71}{\, \micro\meter/pixel}, respectively.

The jetting procedure followed in the present work is identical to the one in our previous work on drying of binary liquid mixtures in an inkjet printhead~\cite{RumpDrying}. The experiments always started by jetting 999 droplets at a drop-on-demand (DoD) frequency of \SI{1}{\, \kilo\hertz} to ensure that the liquid mixture in the nozzle had the same composition as in the bulk. Then, the jetting was temporarily ceased for the predetermined idle time, which is a control parameter in the experiments. When the desired idle time had passed, the piezo was driven to produce a single droplet that was imaged using either stroboscopic or high-speed imaging. The driving voltage of the piezo $V_{\text{max}}$ as well as the pulse width $t_w$ (Fig.~\ref{fig:ExpSetups}c) were varied to control the amount of liquid jetted from the nozzle.

\begin{figure*}[htb]
	\centering
	\includegraphics[width=1\textwidth]{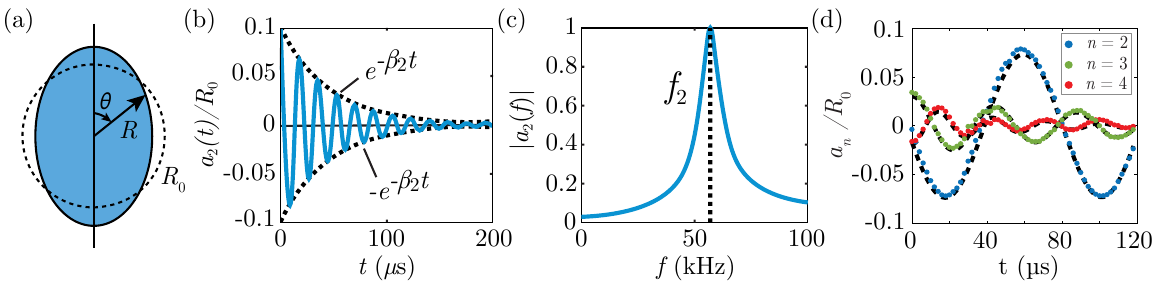}
	\caption{(a) Schematic showing the droplet during axisymmetric oscillations around the vertical axis, where the dashed line indicates the equilibrium droplet shape. (b) Temporatl variation of the normalized oscillation amplitude of mode $n = 2$. The damping rate $\beta$ is determined from the decay envelope (dashed lines) of the amplitude. (c) The frequency of the shape oscillations is determined from the Fourier spectrum of the oscillations. (d) Experimentally measured oscillation modes $n = 2$ (blue markers), $n = 3$ (green markers), and $n = 4$ (red markers) 	for the 1 CMC Dynol solution. The dashed lines show the best fit of Eq.~\eqref{Eq:AMode}, which thereby provides the surface tension $\gamma$.}
	\label{fig:schematicSurftens}
\end{figure*}

\subsection{Image analysis}
The acquired images were analyzed in \textsc{Matlab}. First, the position and velocity of the liquid exiting the nozzle were extracted by tracing the front of the ejected liquid along the direction of jetting. This was followed by a moving average smoothing filter and subsequently, the derivative of the smoothed position data was calculated to find the velocity. The time between the start of the actuation of the piezo and pinch-off was documented as the pinch-off time. The error in the pinch-off time was set to the interframe time of the experiment. After being ejected from the nozzle, the liquid filament forms one or multiple droplets. These droplets were in the field of view for at least another \SI{50}{\micro\second}. By recording the droplet in flight, the mean radius and its standard deviation were calculated from at least three independent experimental realizations. A linear fit to the position data and the fitting error provided the velocity of the droplet and its error, respectively. As the total volume of liquid jetted out of the nozzle was not constant from droplet-to-droplet, we quantified the total momentum of the droplets, defined as the sum of the mass of all the jetted droplets multiplied by their individual velocities. Dividing the momentum by the total mass then gives a mass-averaged velocity.

Finally, the frequency of the oscillating droplet was extracted to calculate the surface tension. This method~\cite{Staat2017,Yang2014,Bzdek2020,Hoath2015} is based on Rayleigh's expression for the shape of an axisymmetrically deformed droplet as a sum of Legendre polynomials $P_n$~\cite{Rayleigh1879}:
\begin{equation}\label{Eq:legendre}
R(\theta,t) = \sum_{n=0}^{\infty}a_n(t)P_n(\cos(\theta)),
\end{equation}
where $\theta$ is the polar angle and $a_n(t)$ is the time-dependent surface mode coefficient for mode number $n$. Assuming an incompressible liquid, no evaporation, and no internal rotation, $a_0(t) = R_0$, where $R_0$ is the radius of the unperturbed spherical droplet (see Fig.~\ref{fig:schematicSurftens}a). By having the origin of the polar coordinate system at the center of mass of the droplet, the modal coefficient of the next mode becomes zero ($a_1(t) = 0$). Thereby, Eq.~\eqref{Eq:legendre} simplifies to:
\begin{equation}
R(\theta,t) = R_0 + \sum_{n=2}^{\infty}a_n(t)P_n(\cos(\theta)).
\end{equation}
The modal coefficients themselves can be written as damped oscillations:
\begin{equation}\label{Eq:AMode}
a_n(t) \sim e^{-\beta_n t}\cos(\sqrt{\left(\omega_n^2-\beta_n^2\right)} t),
\end{equation}
where $\beta_n$ is the damping rate and $\omega_n$ the angular frequency for each mode $n$. The extraction of the damping rate and eigenfrequency using Eq.~\eqref{Eq:AMode} is demonstrated in Figs.~\ref{fig:schematicSurftens}b and c, respectively. The eigenfrequency can be related to the surface tension by:
\begin{equation}
\omega_n^2 = (2\pi f_n)^2 = n (n-1)(n+2)\frac{\gamma}{\rho R_0^3},
\end{equation}
where $f_n$ is the eigenfrequency, $\gamma$ the surface tension, and $\rho$ the density. Also, the bulk viscosity $\mu$ can be extracted from the damping rate of the droplet oscillations:
\begin{equation}
\beta_n = (n-1)(2n+1)\frac{\mu}{\rho R_0^2}.
\end{equation}
Note that this model is for a Newtonian liquid, and that (potential) effects of surface elasticity and viscosity are not taken into account. More complex models including surface rheology are available~\cite{Tian1995}. However, here, the simple Newtonian model is only employed to conclude on the timescale at which the surfactants start to affect the apparent surface tension rather than on the obtained absolute values, which may indeed be inaccurate in the strictest quantitative sense. 
From the experimental measurements, the modes $n = 2$ through $n = 4$ were extracted ($n > 5$ decay very fast), of which a typical example is shown in Fig.~\ref{fig:schematicSurftens}d.


\subsection{Bulk viscosity measurement}
The bulk viscosities of the aqueous Triton solutions with concentrations above 1 CMC were measured by a capillary viscometer~\cite{Pham2018}. The solution was pushed through a 2.3 cm long and 150 \si{\micro\meter} diameter capillary tube at a pressure of \SI{1.5}{\bar}. Knowing the flow rate $Q$ and the pressure drop $\Delta P$ across the tube, the viscosity $\mu$ can be calculated using the Hagen-Poiseuille equation~\cite{book-kundu}:
\begin{equation}
\mu = \frac{\Delta P \pi r^4}{8 L Q},
\end{equation}
where $r$ is the radius of the tube of length $L$. The flow rate is determined by weighing the amount of fluid that has passed through the tube in \SI{60}{\, \second}, and then dividing it by the density $\rho$.

\section{Numerical model}\label{sec:num}

A numerical model was used to simulate the spatiotemporal distribution of Triton within the nozzle and during the droplet formation process as this concentration field cannot be directly obtained from the experiments. We limit ourselves to solving only the mass transport equations of Triton in the bulk liquid, i.e. without any adsorption kinetics, as incorporating surfactant material properties requires knowledge about adsorption and desorption rates when the Triton concentration exceeds the CMC value.\\

Simulating the entire process, i.e. the preferential evaporation of water from the nozzle, subsequently followed by a jetting event, is a challenging problem, which we have addressed in our previous work~\cite{RumpDrying}. In this section, we summarize the salient features of that work and describe the necessary alterations that were required as a different printhead was used.\\

As a general framework, we have used a sharp-interface arbitrary Lagrangian-Eulerian finite element method (ALE-FEM) expressed in axisymmetric cylindrical coordinates. This comes with the benefit that the liquid-gas interfaces are always exactly represented by sharp curves, which easily allows for incorporating Marangoni flow and mass transfer. Furthermore, FEM is solved implicitly via Newton's method, which provides a stable solution method along with flexible time stepping on the two different timescales of evaporation and jetting. The present implementation is based on the finite element library \textsc{oomph-lib}~\cite{oomphlib,Heil2006}.

\subsection{Evaporation phase}
During the evaporation phase, the fluid dynamical equations in both the gas and the liquid phase were solved. Since there is no actuation, all parts of the driving were deactivated. The governing equations for solving the evaporation phase resemble the equations which have been used successfully in previous works on the evaporation of multi-component droplets on substrates, see e.g. reference~\cite{Li2020}.


\subsubsection{Vapor diffusion in the gas phase}
We assume diffusion-limited evaporation, i.e. the evaporation rate of water can be obtained by solving the vapor diffusion equation for the partial mass density $\VapConcW$ of water vapor in the gas phase:
\begin{align}
\partial_t\VapConcW= D_\text{vap} \nabla^2 \VapConcW,
\end{align}
subject to the boundary conditions:
\begin{align}
\VapConcW&=\VapConcWVLE(\MassFracW) &\text{at the liquid-gas interface} \,,\\
\VapConcW&=\VapConcWInfty & \text{far away}\,,
\end{align}
i.e. at the liquid-gas interface, we impose the vapor-liquid equilibrium according to Raoult's law,  $\VapConcWVLE(\MassFracW)=x(w)p_\mathrm{sat}^\star M/(RT)$, where the mole fraction of water $x$ is calculated from the water mass fraction $\MassFracW$, and $p_\mathrm{sat}^\star$, $M$, $R$, and  $T$, are the vapor pressure of pure water, its molar mass, the universal gas constant, and the temperature, respectively.
The ambient water vapor concentration $\VapConcWInfty$ far away is not directly imposed at the distant boundaries of the considered gas domain, since it would induce considerable errors stemming from the finite size of the considered gas mesh. Instead, a Robin boundary condition mimicking an infinite domain is used, which is based on a multipole expansion truncated at monopole order~\cite{Diddens2017}. In comparison to water, the volatility of Triton is negligible; therefore we do not account for Triton evaporation here.

\subsubsection{Multi-component flow in the liquid phase}

The bulk flow in the liquid phase is governed by the Navier-Stokes equations with a composition-dependent mass density $\rho$ and viscosity $\mu$ together with the advection-diffusion equation for the water mass fraction field $\MassFracW$:
\begin{align}
\rho\left(\partial_t\mathbf{u}+\mathbf{u}\cdot\nabla\mathbf{u}\right) &\notag  \\  = -\nabla p + \nabla\cdot\left[\mu\left(\nabla\mathbf{u}+(\nabla\mathbf{u})^\mathrm{T}\right)\right] \label{eq:FEM:navstokes} ,\\ 
\partial_t \rho + \nabla\cdot\left(\rho\mathbf{u}\right)&=0 \label{eq:FEM:contieq},\\
\rho\left(\partial_t \MassFracW + \mathbf{u}\cdot\nabla \MassFracW\right)=\nabla\cdot\left(\rho D\nabla\MassFracW\right). 
\label{eq:FEM:compoadvdiff}
\end{align}
The composition-dependent properties $\rho(\MassFracW)$ and $\mu(\MassFracW)$ are described as linearly dependent on the composition. 
As mentioned in the introduction, equation of state models for surface tension are only valid up to the CMC. Therefore, we chose to fit the following equation to experimental data~\cite{Ullah2019}:
The surface tension $\sigma(\MassFracW)$ is described as the function $A-B/(1+C (1-\MassFracW)^2)$, where the coefficients are set as $A$ = 33~mN/m, $B$ = -39.5~mN/m, and $C = \num{36e6}$. The result is a curve that reaches the CMC concentration with a surface tension of \SI{37}{\milli\newton/\meter}, from which it slowly decreases to the value of \SI{33}{\milli\newton/\meter}. 
The diffusivity $D$ is fixed at a value of \SI{8e-11}{\meter\squared/\second}~\cite{Venditti2022}.
At the liquid-gas interface, normal and tangential stress balances, i.e. Laplace pressure and Marangoni shear, are applied without consideration of the stresses in the gas phase, which can be disregarded due to the small density and viscosity ratios:
\begin{align}
\mathbf{n}\cdot\mathbf{T}\cdot\mathbf{n}&=\kappa\sigma \, , \label{eq:FEM:laplpress}\\ 
\mathbf{n}\cdot\mathbf{T}\cdot\mathbf{t}&=\nabla_\text{S}\sigma \label{eq:FEM:marastress} \, ,
\end{align}
with the stress tensor $\mathbf{T}=-p\mathbf{1}+\mu(\nabla\mathbf{u}+(\nabla\mathbf{u})^\mathrm{T})$, and the normal and tangential unit vectors $\mathbf{n}$ and $\mathbf{t}$, respectively. $\kappa$ is the curvature of the interface and $\nabla_\text{S}$ the surface gradient operator.
The kinematic boundary condition considering water evaporation reads as
\begin{align}
\rho\left(\mathbf{u}-\mathbf{u}_\text{I}\right)\cdot\mathbf{n}=\MassTransW  \label{eq:FEM:kinbc},
\end{align}
which connects the normal liquid bulk velocity $\mathbf{u}$ with the normal interface velocity $\mathbf{u}_\text{I}$ via the evaporation rate $\MassTransW=-D_\text{vap}\nabla c\cdot\mathbf{n}$. The liquid-solid interfaces within the simulated printhead geometry are no-slip and no-flux boundaries, i.e. ($\mathbf{u}=0$ and $\nabla w\cdot\mathbf{n}=0$). At the top of the simulation domain, we also use $\partial_z w=0$, since the diffusion profile does not reach this boundary in the evaporation times considered.
Finally, the evaporation of water leads to a change of the liquid composition near the interface, which is incorporated via the boundary condition
\begin{align}
-\rho D\nabla\MassFracW\cdot\mathbf{n}=(1-\MassFracW)\MassTransW\, . \label{eq:FEM:evapcompochange}
\end{align}
The far field in the reservoir is again mimicking an infinite domain by a far-field Robin boundary condition. Furthermore, a constant underpressure of \SI{8}{\, \milli\bar} is applied as in the experiments, which results in a slightly inward curved meniscus.
As initial conditions, a vanishing velocity field, the underpressure -\SI{8}{\, \milli\bar}, and a homogeneous composition is used. The initial meniscus shape is in equilibrium with the underpressure and the gas phase is initialized with $c(t{=}0)=c_\infty$.

Since we are mainly interested in a qualitative profile of the bulk composition in the droplet during jetting, we have disregarded the influence of evaporative cooling, i.e. we consider an isothermal setting for simplicity. While isolated small volumes of water can cool down considerably due to evaporation, i.e. by a few to several kelvins even in short times \cite{Wakata2024}, in reality, e.g. in a MEMS printhead, the liquid is surrounded by a solid that ensures sufficient heat conduction. The thermal Marangoni effect, that could potentially influence the composition distribution, can be disregarded, since even tiny amounts of surfactants can entirely arrest thermal Marangoni flows during evaporation \cite{vanGaalen2022}.

\begin{figure*}[htb]
	\includegraphics[width=0.8\textwidth]{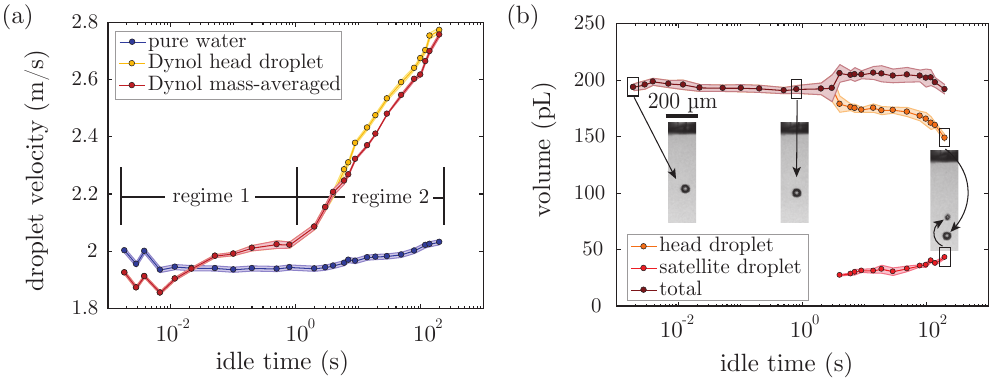}
	\caption{(a) Velocities of the pure water droplets and the Dynol head droplets compared to the Dynol mass-averaged velocity (calculated from the momentum) as a function of idle time. (b) Volume of the 1 CMC Dynol droplets and total ejected volume from a \SI{50}{\micro\meter} diameter nozzle, both as a function of idle time. The datapoint markers represent the mean while the shaded regions denote $\pm$ one standard deviation from at least three independent experimental realizations.}
	\label{fig:Dynol2_VolMomVel}
\end{figure*}

\subsection{Droplet formation}
When it comes to the droplet formation, the relevant timescales are several orders of magnitude smaller (faster) than those during the drying. The consideration of the entire system is hence not required anymore. Given the short timescales and the fast convection velocities during jetting, the diffusion-limited evaporation model as used during the evaporation phase is also questionable in this stage. Therefore, the gas phase, and with it the evaporation dynamics, are disregarded during the jetting process. The friction of the jetted droplet in the gas phase is in general not entirely negligible, but its influence on the drop formation is. This has been shown by the excellent agreement between experiments and frictionless numerics in the slender jet approximation~\cite{vdBos2014}.
In the present ALE-FEM simulations, the deactivation of the gas phase and evaporation implies setting $\MassTransW=0$ in Eqs.~\eqref{eq:FEM:kinbc} and \eqref{eq:FEM:evapcompochange}.

\subsubsection{Modeling of the piezo dynamics}
To bring the liquid into motion, at the top of the nozzle, the normal traction is set to:
\begin{equation}
\mathbf{n}\cdot\mathbf{T}\cdot\mathbf{n}= -\alpha V_{pulse}(t),
\end{equation}
where $\mathbf{n}$ is the normal, $\mathbf{T}$ the stress tensor ($\mathbf{T}=-p\mathbf{1}+\mu(\nabla\mathbf{u}+(\nabla\mathbf{u})^\mathrm{T})$), $V_{pulse}(t)$ the pulse shape and amplitude, and $\alpha$ the pulse-to-pressure conversion factor, which is set to \SI{400}{\pascal/\volt}.

\subsubsection{Sharp-interface ALE method with topological changes}
While the sharp-interface ALE method has the benefits of easily and accurately incorporating Marangoni flow and evaporation, one of its major drawbacks compared to e.g. volume-of-fluid or phase-field models is the treatment of topological changes, i.e. the pinch-off of droplets from the jet or their potential subsequent in-air coalescence.
For simple axisymmetric problems, however, these events can be treated by mesh reconstruction: after each accepted time step, the liquid-gas interface is tested for parts that run nearly parallel to the axis of symmetry. If these are close to the axis (i.e. within \SI{2}{\percent} of the nozzle radius), the pinch-off position is estimated by finding the thinnest spot of the tail that also shows a profile of local relative outflux, i.e. a relative velocity that changes sign in the vicinity. Whenever such a position can be found, the liquid-gas mesh is artificially split and reconnected to the axis of symmetry. Afterwards, a new separated mesh is constructed and all relevant fields are interpolated from the previous mesh, whereby the data of the nodes at the liquid-gas interface are interpolated from the data stemming from the previous still-connected interface only.

While this method introduces an artificial length scale, i.e. the thickness threshold for pinch-off to occur, all other numerical methods also intrinsically have these artificial scales, be it the cell size in a volume-of-fluid approach, the interface thickness in a phase field approach, or the regularization radius in the slender jet (lubrication theory) method~\cite{Driessen2011}. The method used here for topological changes also showed perfect agreement with experiments on droplets colliding mid-air~\cite{Hack2021b}.

\section{Results \& discussion} \label{S4}

We start our investigation on the role of surfactants on droplet velocity by measuring the droplet velocities of pure water and that of an aqueous 1~CMC Dynol solution for idle times spanning over 5 orders of magnitude (from milliseconds to minutes, see Fig.~\ref{fig:Dynol2_VolMomVel}a). The trends in the velocity of the surfactant solution can be divided into two regimes: the minor difference compared to pure water during shorter idle times (less than \SI{1}{\second}; \emph{regime 1}), and a monotonic increase at longer idle times (beyond \SI{1}{\second}; \emph{regime 2}). Remarkably, the velocity increases up to 40\% compared to that of pure water at an idle time of \SI{200}{\, \second}, while the droplet velocity of pure water remained relatively constant as the idle time increased, i.e. it did not increase by more than 3\%. Figure~\ref{fig:Dynol2_VolMomVel}b shows that the total jetted volume of the surfactant solution increased when the droplet velocity also increased for the corresponding idle times. The figure also shows that at idle times longer than \SI{3}{\second}, two droplets were formed, i.e. a head droplet and a satellite droplet.  Therefore, we calculate the sum of the momentum of both droplets (head droplet and satellite) and normalize it by the total mass to obtain the mass-averaged velocity of the jetted liquid, plotted in Fig.~\ref{fig:Dynol2_VolMomVel}a (red data points). The mass-averaged velocity is very similar to the velocity of the head droplet, i.e. the velocity of the satellite droplet was very similar to that of the head droplet. Thus, the velocity increase is not due to a reduction in droplet size, but due to an increase in the kinetic energy of the jetted liquid. Regimes 1 (shorter than \SI{1}{\second}) and 2 (longer than \SI{1}{\second}) are further discussed in the following sections.

\subsection{Regime 1: Short timescales ($<$ \SI{1}{\second})}

\begin{figure*}[htb]
	\centering
	\includegraphics[width=.7\textwidth]{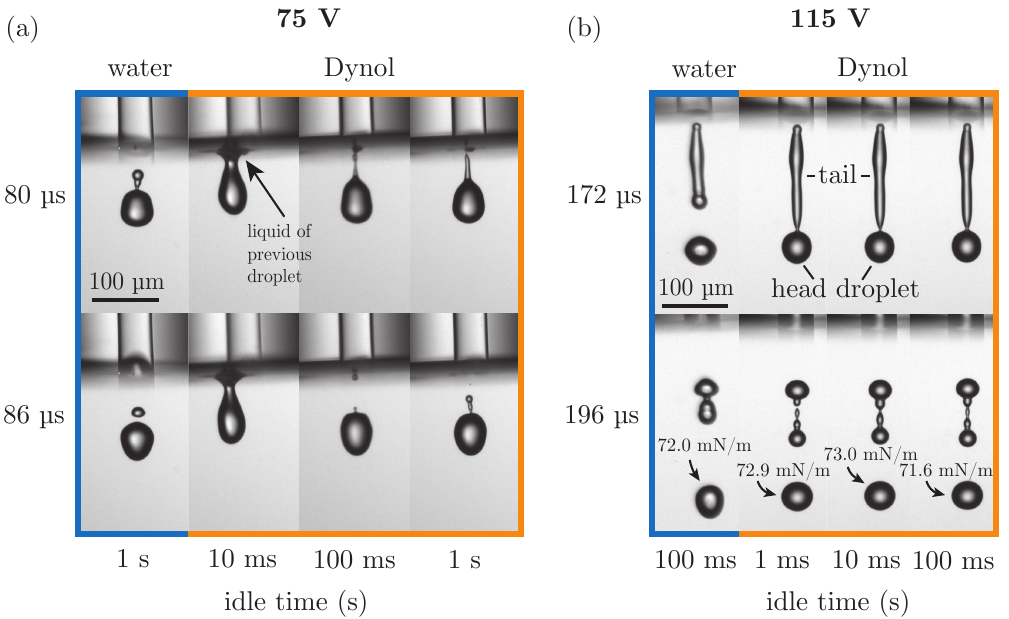}
	\caption{Droplet formation for water and an aqueous 1~CMC Dynol solution at different idle times. (a) At \SI{75}{\volt}, the top row compares the experiments at \SI{80}{\micro\second}, just before the surfactant solution filament pinches off from the nozzle plate, and the bottom row compares the experiments at \SI{86}{\micro\second}, just before the filament of the 1 s surfactant solution breaks up. (b) At \SI{115}{\volt}, the top row compares the experiments at \SI{172}{\micro\second}, just before the surfactant solution filament breaks up, while the bottom row compares the experiments at \SI{196}{\micro\second}, just before the tail droplet for \SI{100}{\milli\second} breaks up. The bottom row also indicates the surface tension coefficients of the head droplet as calculated from the droplet oscillations. }
	\label{fig:BreakDynol}
\end{figure*}

We attribute the minor difference in droplet velocity between pure water and the surfactant solution in regime~1 in Fig.~\ref{fig:Dynol2_VolMomVel}a to wetting effects. This is demonstrated in Fig.~\ref{fig:BreakDynol}a where, at the shortest idle times, a droplet of surfactant solution left on the nozzle plate from the preceding droplet formation process influences the current droplet. 

The droplet formation processes in Fig.~\ref{fig:BreakDynol}a show the delay in pinch-off for the surfactant solution and the filament break-up. However, the higher voltage (115 V) case in Fig.~\ref{fig:BreakDynol}b shows that, next to the similar delay in filament break-up, the pinch-off from the nozzle for the surfactant solution is not dissimilar to the pure water droplet. This is also as described in reference \cite{Antonopoulou2021}, which for a surfactant solution reports a delayed filament break-up but an unaffected pinch-off from the nozzle. That paper shows visualizations from numerics revealing that the surfactant is  transported toward the trailing edge of the head droplet during the droplet formation process. The resulting gradient in surfactant concentration, and correspondingly in surface tension, leads to a Marangoni flow, which causes the observed delay in filament break-up.

Figures~\ref{fig:BreakDynol}a and b also demonstrate the minor influence of the idle time on the filament break-up between the surfactant solutions. The measurement of the droplet shape oscillations provides the surface tension of the head droplets in Fig.~\ref{fig:BreakDynol}b, which were all between 72 and 74~\SI{}{\milli\newton/\meter}, including that of the water droplet. Even though the meniscus at the nozzle exit accounts for only $\approx$~25\% of the surface area of the head droplet, this was expected to result in a decrease in surface tension with increasing idle time. The lack of a decrease of surface tension suggests that the surfactant initially present on the meniscus is rapidly transported across the interface, ending up being either distributed along the entire interface of the jetted filament or accumulating in its tail. Further  investigations are required to understand this observation, which is beyond the scope of the present work.

\subsection{Regime 2: Long timescales  ($>$ \SI{1}{\second})}

\subsubsection{Droplet formation}

\begin{figure*}[htb]
	\centering
	\includegraphics[width=\textwidth]{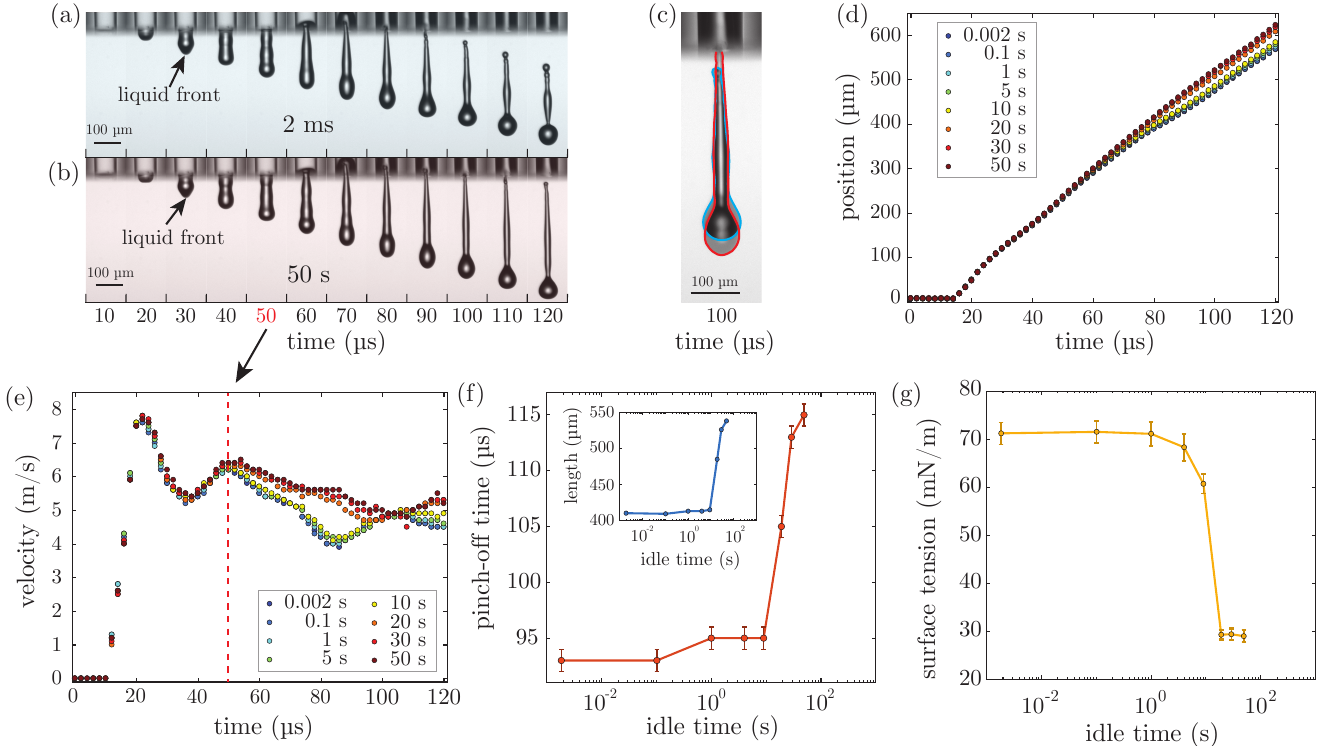}
	\caption{Time-lapsed snapshots of droplet formation of 1 CMC Dynol from a \SI{70}{\micro\meter} nozzle diameter for an idle time of (a) \SI{2}{\, \milli\second} and (b) \SI{50}{\second}. (c) Overlap of the droplet shapes at \SI{100}{\, \micro\second} from the experiments: \SI{2}{\, \milli\second} idle time in red and \SI{50}{\, \second} idle time in blue, showing clear differences. (d) The position of the liquid front for multiple idle times. (e) The velocity of the liquid front for multiple idle times. (f) The pinch-off time of the liquid column from the nozzle for multiple idle times; inset shows the length of the liquid column just after pinch-off from the nozzle for multiple idle times. (g) The surface tension measured from the head droplet's oscillations for multiple idle times. The datapoint markers in (f) and (g) represent the mean while the error bars denote $\pm$ one standard deviation from at least three independent experimental realizations.}
	\label{fig:Dynol_tracking}
\end{figure*}

To investigate the increase of the ejection velocity at long timescales in regime 2, we first compare the droplet formations as measured for the 1~CMC Dynol solution at two idle times in Figs.~\ref{fig:Dynol_tracking}a and b, where (a) corresponds to an idle time of \SI{2}{\milli\second} and (b) to an idle time of \SI{50}{\second}. The most striking difference is the length of the tail. This is highlighted in Fig.~\ref{fig:Dynol_tracking}c, which shows that the increase in tail length is due to both the front of the liquid filament having traveled further (higher velocity) and the end of the tail pinching off later in time. We plot the front position of the filament vs. time in Fig.~\ref{fig:Dynol_tracking}d for idle times ranging from \SI{2}{\, \milli\second} up to \SI{50}{\, \second}. The velocity calculated from these position data is shown in Fig.~\ref{fig:Dynol_tracking}e. Note that the velocity curves start to deviate only after \SI{50}{\, \micro\second} from the start of the piezo actuation pulse, i.e. when the meniscus starts to retract into the nozzle, as shown in Figs.~\ref{fig:Dynol_tracking}a and b. For idle times longer than $\approx$\SI{10}{\second}, the velocity of the filament decreases less during the retraction of the meniscus (Fig.~\ref{fig:Dynol_tracking}c). This observation suggests that the surface tension at the rear of the filament is decreased at longer idle times, thereby exerting a lower pulling-force on the filament during meniscus retraction. The presence of surfactants near the nozzle exit is further indicated by the delay in pinch-off time, as indicated in Fig.~\ref{fig:Dynol_tracking}f. The increase in pinch-off time can be caused by a decreased surface tension but also by an increased surface viscosity due to the presence of the Dynol surfactant~\cite{Kamat2018,Wee2021}. Moreover, the measured surface tension of the head droplet in Fig.~\ref{fig:Dynol_tracking}g demonstrates a correlation between the increase in droplet velocity and a decrease in surface tension. This indicates that at longer idle times, the surfactants are not only present near the rear of the liquid filament, but also at the front. However, the increase in droplet velocity is not due to a change in the initial velocity, but due to the changes at the rear of the droplet: the ejected liquid is not slowed down as much by the retracting meniscus and both the filament length and pinch-off time increase.

\subsubsection{Evaporation}

\begin{figure*}[htb]
	\centering
	\includegraphics[width=.8\textwidth]{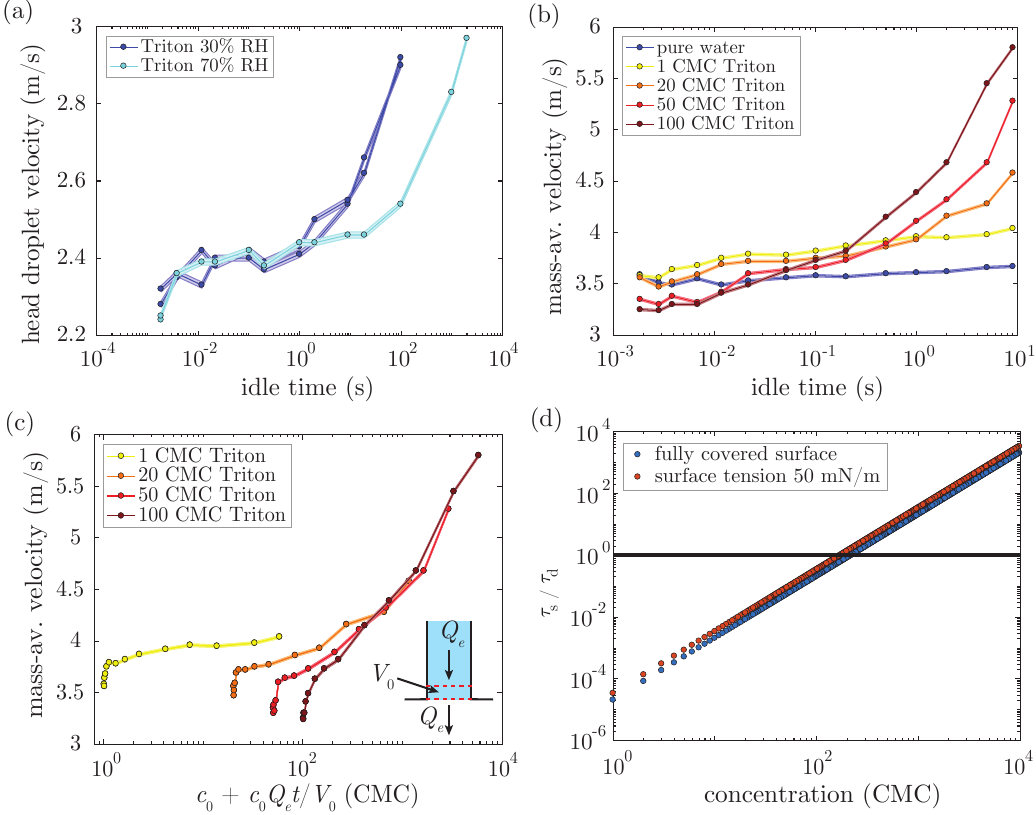}
	\caption{(a) Head droplet velocity for a 1~CMC Triton solution from a \SI{50}{\,\micro\meter} diameter nozzle at two different relative humidities as a function of idle time. The datapoint markers represent the mean while the shaded area denotes $\pm$ one standard deviation of at least three independent experimental realizations. (b) Mass-averaged (mass-av.) velocities for pure water and 1--100~CMC Triton solutions from a \SI{50}{\, \micro\meter} diameter nozzle as a function of idle time. (c) Mass-averaged velocities for the different Triton concentrations as a function of the estimated Triton concentration in the control volume $V_0$ at the nozzle exit (sketched in the inset as the red dotted lines). (d) Ratio of the timescales for the droplet surface formation $\tau_S$ and the Triton diffusion $\tau_D$, for a fully covered surface (blue) and a surface coverage corresponding to \SI{50}{\, \milli\newton/\meter}~\cite{Horozov2000} (red).}
	\label{fig:VeloRHConc}
\end{figure*}

The combination of a higher surfactant concentration at the surface and the requirement of long timescales suggests that selective evaporation at the meniscus generates this velocity increase. To be able to perform experiments with bulk surfactant concentrations above the CMC, in what follows, we use Triton surfactant as the Dynol surfactant phase separates (oils out) at concentrations higher than the CMC~\cite{Tag2010}. On the other hand, Triton forms nanometric micelles at bulk concentrations above the CMC~\cite{Paradies1980}. We first confirmed that the observed increase in droplet velocity with idle time is also present for jetting experiments with a 1~CMC Triton solution, see Fig.~\ref{fig:VeloRHConc}a. Indeed, a similar trend is observed: the droplet velocity increases with idle time. However, significant differences are present as well. First, the increase of the head droplet velocity at an idle time of \SI{200}{\second} and a relative humidity of 70\% was only 6\% (Fig.~\ref{fig:Dynol2_VolMomVel}a), whereas for Dynol the corresponding increase was 40\% (Fig.~\ref{fig:Dynol2_VolMomVel}a). Moreover, the time-dependent velocity of the liquid filament during its ejection from the nozzle was different. For Triton, the increase in droplet velocity appears to be present quite early on in the ejection process (see supplementary information Fig.~S3), whereas for Dynol it was only apparent \SI{50}{\, \micro\second} after the start of the piezo actuation pulse (Fig.~\ref{fig:VeloRHConc}e). However, and more importantly, an increasing trend of the velocity with idle time is still observed. On top of that, the curve in Fig.~\ref{fig:VeloRHConc}a measured at the decreased relative humidity of 30\% shows that the sharp increase in droplet velocity occurs at a shorter idle time as compared to the 70\% relative humidity case. This confirms that evaporation is the driving parameter for the increase of the droplet velocity with idle time.

\subsubsection{Increased local concentration}
The observed influence of evaporation suggests that a surfactant concentration higher than 1~CMC is required at the nozzle exit to increase the droplet velocity. Therefore, the measurements were repeated with bulk Triton concentrations of 20, 50, and 100~CMC. Increasing the Triton concentration above 100~CMC made the solution too viscous to jet at the applied amplitude of the piezo driving pulse. The mass-averaged droplet velocity obtained for the different Triton concentrations is plotted in Fig.~\ref{fig:VeloRHConc}b as a function of the idle time. At short timescales ($<$~\SI{0.1}{\second}), the increased Triton concentration resulted in a decrease in the droplet velocity instead of the expected increase. The decrease in velocity most likely results from the increased bulk viscosity of the high-concentration Triton mixtures (see supplementary information Fig.~S4). At longer idle times ($>$~\SI{0.1}{\second}), however, the increase in droplet velocity occurred earlier for higher Triton concentrations. This observation suggests that the required increase in local concentration in the nozzle for the same increase in droplet velocity is reached earlier for higher initial bulk concentrations.

We try to estimate the required increase in local surfactant concentration in the nozzle by considering a control volume at the nozzle exit (red dashed lines in inset of Fig.~\ref{fig:VeloRHConc}c), out of which the water evaporates and is replaced by surfactant solution from the bulk. This means that we assume the diffusivity of the surfactant to be negligible, for simplicity. The local concentration $c$ of the surfactant in this volume $V_0$ is increasing with idle time $t$ beyond the initial concentration $c_0$ due to the evaporative flux $Q_e$, given by: $c(t) =c_0 + c_0 Q_e t/V_0$. The evaporative flux was estimated from the numerical simulations to be \SI{0.02}{\kilogram/(\meter\squared\second)}. The control volume covers the nozzle exit surface area and we choose a depth of \SI{1}{\, \micro\meter}. Figure~\ref{fig:VeloRHConc}c shows that the droplet velocity curves now collapse at local surfactant concentrations $c(t)$ beyond $\approx$~400~CMC. Increasing the size of the control volume still collapses the data, but at a longer idle time. Even though this concentration estimation is based on a simple calculation and the absolute values should not be trusted, the collapse of the data provides us with the valuable insight that the local surfactant concentration in the nozzle should be orders of magnitude higher than the CMC to increase the droplet velocity. For example, the 20~CMC solution contains $\approx$~35 times the initial surfactant concentration when the droplet velocity starts to increase (after an idle time of \SI{5}{\, \second}). These results suggest that a thin layer of highly concentrated surfactants at the nozzle exit is required to cause a significant observable increase in droplet velocity.

To investigate how this high concentration of surfactant can alter the jetting
behavior, we investigate how the adsorption timescale $\tau_D$ (Eq.~\eqref{Eq:tauD}) depends on concentration. First, we note that the adsorption of Triton X-100 is diffusion-limited~\cite{Janczuk1995,Horozov2000}.
We compare $\tau_D$ to the timescale for the droplet formation $\tau_S$ (the time of pinch-off), which is \SI{57}{\, \micro\second} for the Triton experiments. Figure~\ref{fig:VeloRHConc}d shows the ratio of the timescale of droplet formation to the timescale of diffusive surfactant transport, i.e. the P\'eclet number, for the case when the surface is fully covered with surfactant ($\Gamma = 3.4\times10^{-6}\si{\, \mole/\meter\squared}$~\cite{Horozov2000}) and for the case when the surface tension is approximately halfway between pure water and pure surfactant at a surface tension of \SI{50}{\milli\newton/\meter} ($\Gamma = 2.7\times10^{-6}\si{\, \mole/\meter\squared}$~\cite{Horozov2000}). The horizontal black line indicates when the ratio $\tau_S/\tau_D =$ 1, above which the diffusion timescale is shorter than the timescale of droplet formation, both at a surface concentration $\approx$ 200~CMC. This is the same order of magnitude as we observed in Fig.~\ref{fig:VeloRHConc}c, which  demonstrates that the change in droplet formation at longer idle times is due to Triton being able to adsorb fast enough onto the newly formed surface at concentrations above $\approx$ 200~CMC, thereby reducing the surface tension and leading to the observed droplet velocity increase.

\subsubsection{Numerically visualized increased local concentration}

\begin{figure*}[htb]
	\centering
	\includegraphics[width=.7\textwidth]{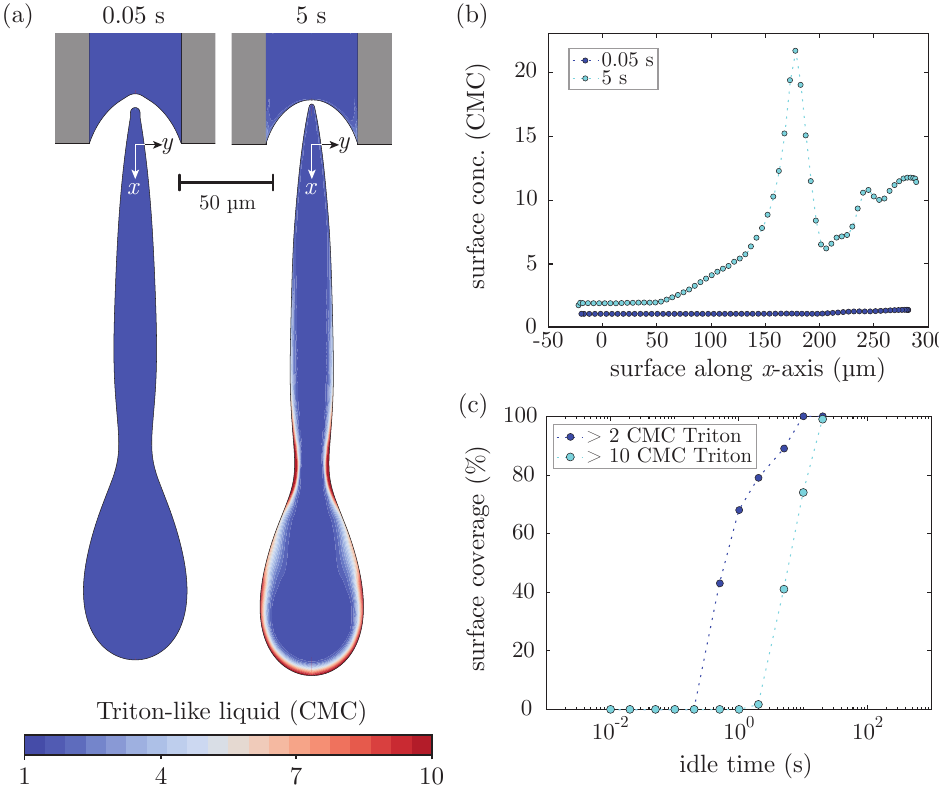}
	\caption{(a) Snapshots at \SI{79}{\, \micro\second} of the numerical simulations of the droplet formation process using a 1~CMC Triton-like solution. The left image shows a snapshot for an idle time of \SI{0.05}{\, \second} and the right image for an idle time of \SI{5}{\, \second}. The colorbar indicates the liquid concentration in a range from 1~CMC to~10 CMC. (b) Surface concentration of the droplets in (a) along the droplet’s axis of symmetry as indicated by $x$. The origin of the $x$-axis is indicated in (a). (c) Percentage of Triton covering the surface at a concentration above 2~CMC and above 10~CMC measured from the simulation data, at the time instant when the droplet has just pinched off from the nozzle, plotted as a function of idle time.}
	\label{fig:Sim}
\end{figure*}

To understand how this thin layer of highly concentrated surfactant solution at the meniscus influences the droplet formation along the entire length of the droplet surface, we turn to numerical simulations. We aim to use these simulations to understand the distribution of the concentration field of Triton during the droplet formation and not its adsorption and desorption kinetics. Therefore, we call it a Triton-like liquid. The numerical methodology used herein has been used before to simulate drying processes in inkjet printing~\cite{RumpDrying}. The resulting droplets for two idle times, \SI{0.05}{\, \second} and \SI{5}{\, \second}, are displayed in Fig.~\ref{fig:Sim}a. The colorbar is kept constant for both images to indicate the differences in concentration of the Triton-like liquid. These two simulations show that after an idle time of \SI{5}{\, \second}, the jetted filament is covered by a thin layer of highly concentrated Triton-like liquid, while at an idle time of \SI{0.05}{\, \second}, the droplet shows a homogenous concentration everywhere. Snapshots corresponding to the droplet formation for \SI{5}{\, \second} can be found in Fig.~S5 of the supplementary information. The wrapping of the surface of the droplet by a high concentration layer of Triton is more convincingly shown in Fig.~\ref{fig:Sim}b, where we plot the concentration of Triton-like liquid along the surface. The surface coverage over time, defined as the percentage of surface area containing the desired concentration compared to the total surface area, for a surface concentration $>$~2~CMC and for a concentration $>$~10~CMC are plotted in Fig.~\ref{fig:Sim}c. These results show that the surfactant surface concentration on the jetted droplet increases beyond the bulk concentration for idle times of the order of seconds -- exactly the timescale at which the experimental droplet velocities were shown to increase (Figs.~\ref{fig:Dynol2_VolMomVel}a and \ref{fig:VeloRHConc}c). Therefore, we hypothesize that the surfactants studied here, which typically adsorb onto a newly formed surface in the millisecond to second timescale, demonstrate a rate of transport onto the surface at the microsecond timescale during droplet formation. This reduction of timescales only happens once the local surfactant concentration at the liquid-air interface of the filament reaches the orders of 100 to 1000 CMC. This presence of surfactants on the surface during the droplet formation reduces the surface tension of the liquid and results in an increase in droplet velocity. We further argue that the increase in drop velocity with idle time for Dynol results from the rear of the filament being covered with surfactant to a larger extent at increased idle times due to a thicker layer of high-surfactant concentration liquid being present at the nozzle exit.

\section{Conclusions and outlook} \label{S5}

In this work, the role of surfactants in inkjet printing was studied at timescales spanning over 6~orders of magnitude, i.e. from the microsecond timescale of the droplet formation process to the timescale of seconds to minutes for the evaporation process taking place between the ejection of two successive droplets. When the time interval between successive droplets was less than one second, surfactants were shown to only influence the details of the break-up dynamics of the jetted filament while the drop velocity and volume remain unchanged from the pure water case. However, droplet velocity and volume increase monotonically with an increase in the time interval between successive droplets when this time interval exceeds one second, i.e. when the droplet production rate is less than \SI{1} droplet per second. The increase in droplet velocity was shown to correlate with a decrease in surface tension. We demonstrate that the observed velocity increase, and surface tension decrease, are evaporation driven. Selective evaporation of water at the nozzle exit for a duration of \SI{10}{\, \second} was estimated to increase the local surfactant concentration by as much as two orders of magnitude. Using numerical simulations, we demonstrate that a thin layer with an increased surfactant concentration located at the nozzle exit can coat the entire jetted liquid filament. We argue that when the surfactant concentration at the nozzle exit reaches a concentration of the order of 100 to 1000 CMC, the adsorption timescale of surfactants can decrease from milliseconds to microseconds, thereby lowering the surface tension and increasing the jet velocity at the microsecond timescale of jet formation in inkjet printing. 

The implications of the unraveling of the thin layer at the nozzle exit covering the entire  surface of the droplet extends beyond the interest of surfactants. This is a property that likely holds true for any material having a local higher concentration at the nozzle exit, including colloidal pigment particles and dyes. Moreover, the results of surfactants having minor influence on the droplet formation at high droplet production frequencies during droplet production in air will be of interest for the development of new ink formulations, as the surfactants are already extensively used for stabilizing colloidal suspensions, as well as the droplet behavior on the substrate.

\section*{Acknowledgements}
This work was supported by an Industrial Partnership Programme of the Netherlands Organisation for Scientific Research (NWO), co-financed by Canon Production Printing Netherlands B. V., University of Twente, and Eindhoven University of Technology. We also acknowledge financial support of the Max Planck Center Twente for Complex Fluid Dynamics. The authors also thank Valeria Garbin for insightful discussions. 


\providecommand{\noopsort}[1]{}\providecommand{\singleletter}[1]{#1}%

\end{document}